\documentclass[prl,twocolumn,superscriptaddress]{revtex4}

\usepackage{graphicx}
\usepackage{amsmath}

\begin{document}

\title{Many-body effects in nonlinear optical responses of 2D layered semiconductors}

\author{Grant Aivazian}
\affiliation{Department of Physics, University of Washington, Seattle, Washington 98195 USA}

\author{Hongyi Yu}
\email[]{yuhongyi@hku.hk}
\affiliation{Department of Physics and Center of Theoretical and Computational Physics, The University of Hong Kong, Hong Kong, China}

\author{Sanfeng Wu}
\affiliation{Department of Physics, University of Washington, Seattle, Washington 98195 USA}

\author{Jiaqiang Yan}
\affiliation{Materials Science and Technology Division, Oak Ridge National Laboratory, Oak Ridge, Tennessee 37831, USA}
\affiliation{Department of Materials Science and Engineering, University of Tennessee, Knoxville, Tennessee 37996, USA}

\author{David G. Mandrus}
\affiliation{Department of Physics and Astronomy, University of Tennessee, Knoxville, Tennessee 37996, USA}
\affiliation{Materials Science and Technology Division, Oak Ridge National Laboratory, Oak Ridge, Tennessee 37831, USA}
\affiliation{Department of Materials Science and Engineering, University of Tennessee, Knoxville, Tennessee 37996, USA}

\author{David Cobden}
\affiliation{Department of Physics, University of Washington, Seattle, Washington 98195 USA}

\author{Wang Yao}
\affiliation{Department of Physics and Center of Theoretical and Computational Physics, The University of Hong Kong, Hong Kong, China}

\author{Xiaodong Xu}
\affiliation{Department of Physics, University of Washington, Seattle, Washington 98195 USA}
\affiliation{Department of Materials Science and Engineering, University of Washington, Seattle, Washington 98195, USA}

\date{\today}

\begin{abstract}
We performed ultrafast degenerate pump-probe spectroscopy on monolayer WSe$_2$ near its exciton resonance. The observed differential reflectance signals exhibit signatures of strong many-body interactions including the exciton-exciton interaction and free carrier induced band gap renormalization. The exciton-exciton interaction results in a resonance blue shift which lasts for the exciton lifetime (several ps), while the band gap renormalization manifests as a resonance red shift with several tens ps lifetime. Our model based on the many-body interactions for the nonlinear optical susceptibility fits well the experimental observations. The power dependence of the spectra shows that with the increase of pump power, the exciton population increases linearly and then saturates, while the free carrier density increases superlinearly, implying that exciton Auger recombination could be the origin of these free carriers. Our model demonstrates a simple but efficient method for quantitatively analyzing the spectra, and indicates the important role of Coulomb interactions in nonlinear optical responses of such 2D materials.
\end{abstract}

\maketitle

Monolayer transition metal dichalcoginides (TMDs), have recently attracted great interests as a new class of two-dimensional (2D) direct band gap semiconductors \cite{KFM_PRL2010,Splendiani_NL2010}. Their band extrema are located at the degenerate but inequivalent $\mathbf K$ and $-\mathbf K$ corners of the hexagonal Brillouin zone. The corresponding gap values are in the visible frequency range which makes them suitable for various optoelectronic applications. Due to the reduced screening in the 2D geometry, the Coulomb interaction effects in monolayer TMDs are remarkably strong. The exciton, the hydrogen-like bound pair of an electron and a hole, then dominates the optical responses of monolayer TMDs. Exceptional excitonic properties in these 2D materials have been revealed, such as large exciton binding energies \cite{Chernikov2014,He2014,Zhang_NL2014,Ye_Nature2014,Ugeda2014,Hanbicki2015,GWang_PRL2015,Zhu_SR2015}, electrostatic tunability between neutral and charged exciton species \cite{Mak2013,Ross2013,Jones_ValleyCoherence}, the interconversion of excitonic valley pseudospin with circularly polarized photons \cite{Xiao2012,Xu_Review2014,Jones_ValleyCoherence,Cao_NatComm2012,Zeng_NatNano2012,Mak_NatNano2012}, and the luminescence upconversion from a charged to a neutral exciton by absorbing an optical phonon \cite{PhotonUpconvert2016}. The strong Coulomb interaction with the free charged carriers also results in a significant band gap renormalization as evidenced by a series of experiments \cite{Ugeda2014,Chernikov2015,Pogna2016,ChernikovPRL2015}.

Nonlinear optical spectroscopies are powerful tools for studying the photo-excitation dynamics with high temporal or frequency resolutions. Unlike PL which only detects the exciton bright states, the signals of nonlinear optical spectroscopy are also sensitive to optically inactive states. The existing experiments in TMDs mainly focus on the exciton population and valley depolarization dynamics, which have revealed the ultrafast lifetimes ($\sim$~ps) of excitons \cite{Mai2014,Sim2013a,Shi_ACSNano2013}, the $\sim$~ps exciton valley depolarization rate \cite{Mai2014,Wang2013,Kumar_Nanoscale2014,Conte_PRB2015}, and the presence of long-lived states \cite{Schaibley_PRL2015}. The physical mechanisms governing the nonlinear optical responses have been well studied in quantum well systems, which includes phase space filling, exciton-exciton interaction, etc. \cite{SuppleMater}. Besides these mechanisms, we also need to emphasize the strong Coulomb effects in monolayer TMDs. One of such effects is the exciton Auger recombination (or exciton-exciton annihilation) process, which appears as an efficient exciton nonradiative decay channel in TMDs \cite{Zhu_SR2015,Kumar_PRB2014,Sun_NL2014,Mouri_PRB2014,Yu_EEA,Poellmann_NM2015}. In carbon nanotubes with even stronger Coulomb effects, such exciton Auger recombination can lead to the generation of free carriers and an induced absorption at the trion resonance \cite{CNT}. It has been shown recently that the band gap renormalization induced by the free carriers can explain the complicated nonlinear spectral shape of TMDs \cite{Chernikov2015,Pogna2016}. Currently the challenge of the nonlinear optical experiments lies in formulating a quantitative model to tease apart the interplay of these different mechanisms.

\begin{figure}[t]
\includegraphics[width=\linewidth]{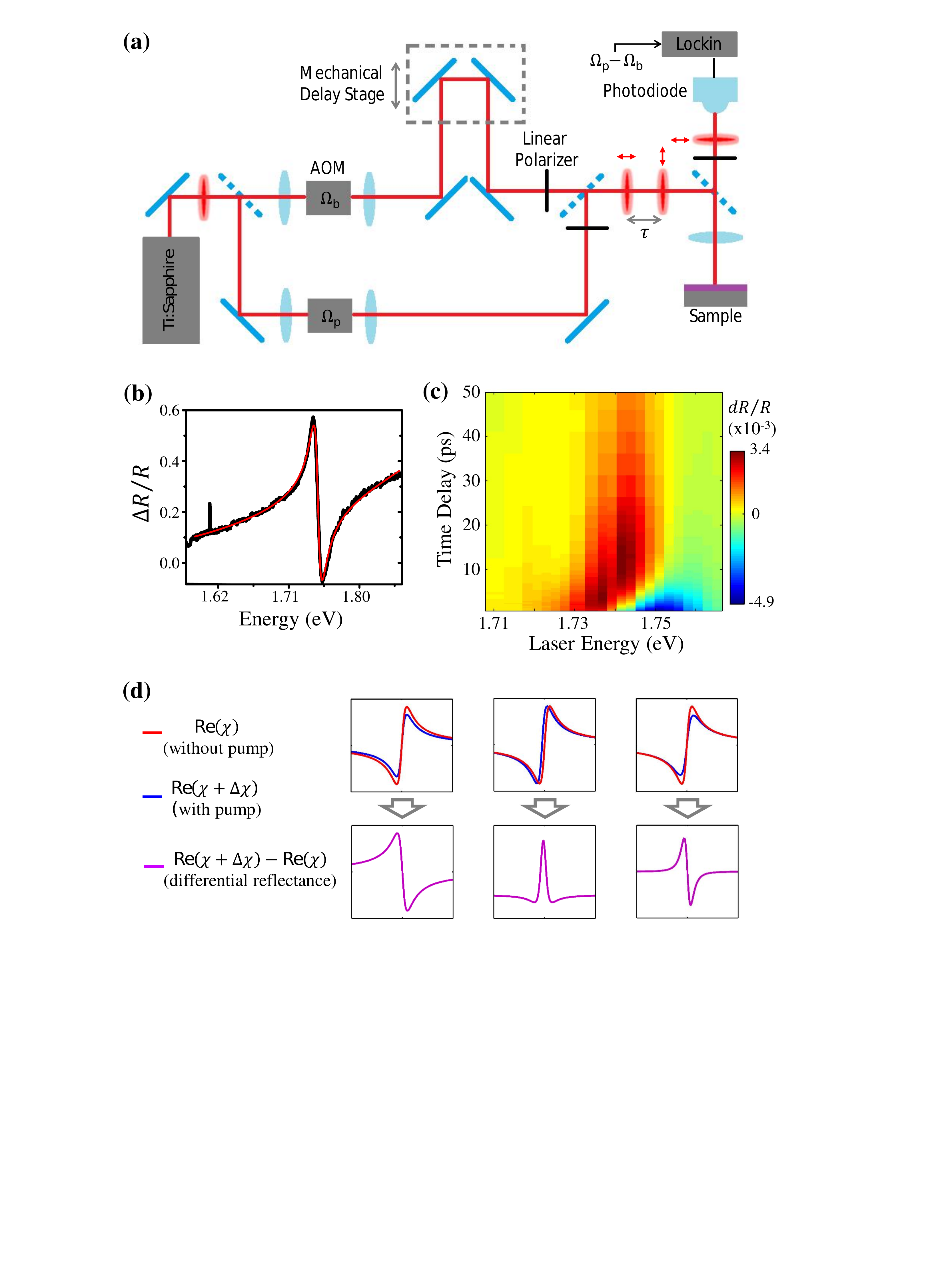}
\caption{\label{fig:one}(a) Schematic of experimental apparatus. (b) White light reflectance contrast $\Delta R/R$ shows resonance near $1.745$~eV. Red line is the fit to an antisymmetric Lorentzian. (c) 2D nonlinear differential reflectance map showing the spectrum time dependence. (d) The resulting nonlinear response due to (from left to right) oscillator strength reduction, resonance shift, and spectral broadening (see Eq. (\ref{eq:NLsusc}) and the associated text). The effect of spectral broadening (right column) is qualitatively similar to the oscillator strength reduction (left column).}
\end{figure}

Here we perform ultrafast degenerate pump-probe spectroscopy in monolayer WSe$_2$ to advance our understanding of the optical dynamics in TMDs and other low-dimensional semiconductors. A pump pulse with frequency near the exciton resonance is applied to generate population excitations in the sample, and we detect the excitation induced change in the reflected probe pulse after a certain time delay. The observed differential reflectance ($dR/R$) spectra exhibit complicated behaviors that can not be simply explained by the excitation induced bleaching. We introduce a simple but efficient and pedagogical model to quantitatively address the nonlinear optical effects and their interplay, which is most suited to extract useful information from large sets of data with complicated spectral shape. A careful analysis to the model reveals the important roles of the many-body interaction from not only the excitons, but also the photo-doped free charged carriers which is characterized by an exciton resonance red shift with a long decay time component (several tens ps) in the spectra response. In contrast, the response from the excited excitons has a short lifetime ($\sim5$~ps) and corresponds to a resonance blue shift. We further analyze power dependence of the $dR/R$ spectra which shows that with the increase of pump power, the exciton population increases linearly and then saturates, while the free carrier density increases superlinearly. This implies that the exciton Auger recombination process could be the origin of these free carriers.

Experiments were performed in a closed-cycle cryostat at $15$~K. The highlights of the experimental apparatus are illustrated in Figure~\ref{fig:one}(a). The output of a mode locked Ti:Sapphire laser ($76$~MHz repetition rate, $\sim0.2$~ps pulse width, $\sim2$~nm FWHM bandwidth) is split into a pump and a probe pulse.  Each beam of pulses is amplitude modulated by individual acousto-optic modulators at frequencies $\Omega_p$ for the pump and $\Omega_b$ for the probe, each around $100$ KHz. The beams are colinearly recombined and focus onto the sample with a high numerical aperture microscope objective ($40$x, NA$=0.65$) to a spot size of roughly $\sim2~\mu$m.  The time delay $\tau$ between the two pulses is controlled by a mechanical delay stage in the pathway of the pump beam.  We measure the pump induced change in reflectance of the probe beam, $dR$, on an amplified photodiode with a lock-in amplifier at the difference modulation frequency $\Omega_p-\Omega_b$.  Since the signal is detected at the difference modulation frequency, only nonlinear effects influenced by both the pump and the probe pulse will be measured. The incident beams' polarizations are set to cross-linear so that a linear polarizer in the detection pathway can be used to filter out the pump beam before detection. The total reflected probe signal $R$ is measured on a separate lock-in amplifier at the probe modulation frequency $\Omega_b$ to normalized the differential reflectance signal $dR/R$. 

Monolayer samples of WSe$_2$ were mechanically exfoliated onto $300$~nm of SiO$_2$ on $n$-doped Si substrates, and were identified through optical contrast. The white light reflectance contrast of the sample, given by $\Delta R/R\equiv(R_\mathrm{SiO_2}-R_\mathrm{WSe_2})/R_\mathrm{SiO_2}$, clearly shows a main feature associated with the A exciton at $E_0=1.745$~eV which can be well fit by an antisymmetric Lorentzian (the background linear offset accounts for the residue influence from the higher energy exciton \cite{Qiu_PRL2013}), see Fig.~\ref{fig:one}(b). 

\begin{figure}[t]
\includegraphics[width=1\linewidth]{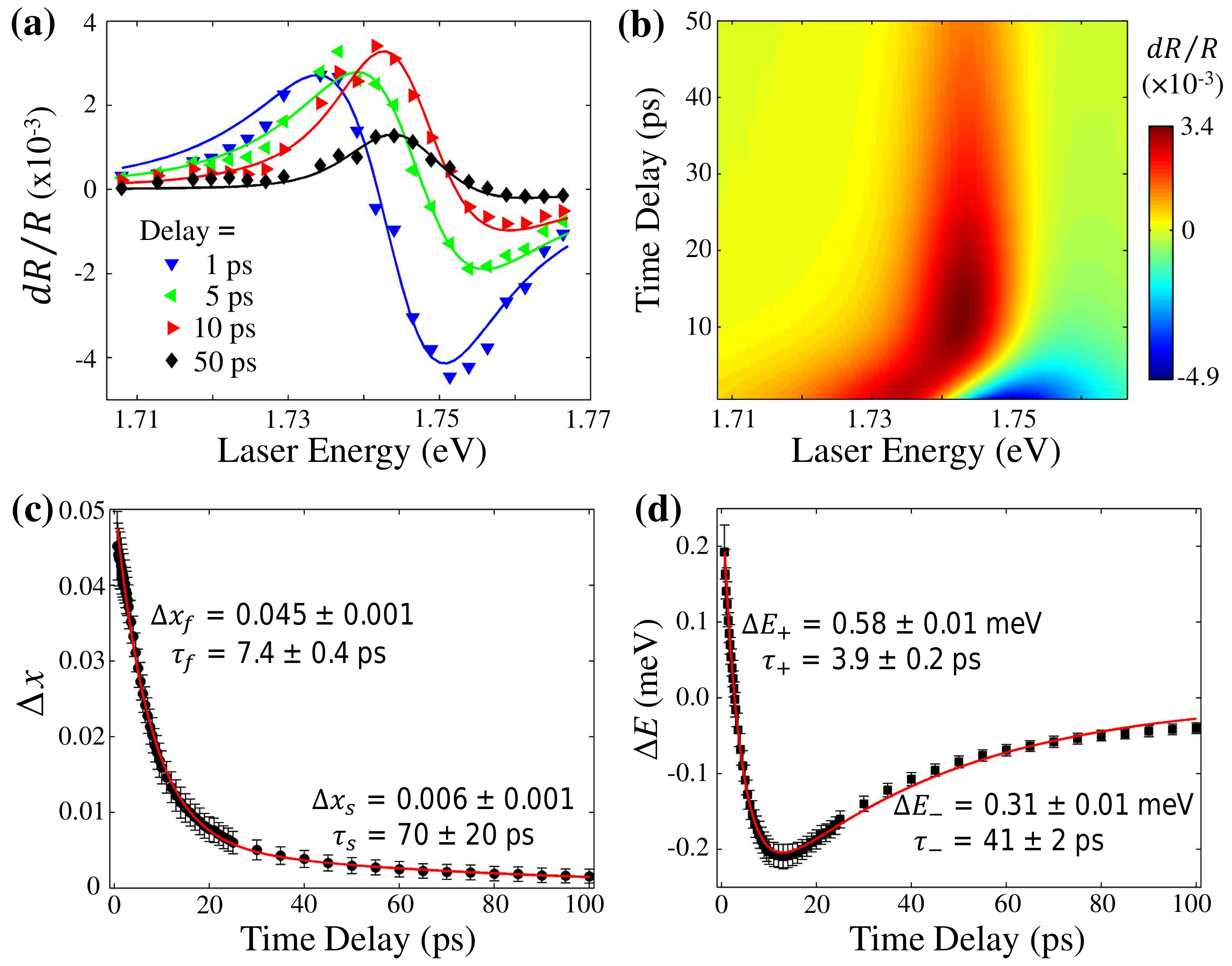}
\caption{\label{fig:two}(a) The measured $dR/R$ spectra (symbols) and the corresponding fits (curves) using Eq. (\ref{eq:NLsusc}) for the given values of time delay $\tau$. (b) The fit result to the 2D data set in Fig. \ref{fig:one}(c). (c) and (d) are plots of the fit parameters $\Delta x(\tau)$ and $\Delta E(\tau)$, respectively, versus the time delay $\tau$. Red curves are lines of best fit to bi-exponential decay functions (Eq. (\ref{eq:biexponential})).}
\end{figure}

For the initial measurement the pump and probe beams have identical powers of $15~\mu$W ($\sim10^{13}$ photons$\cdot$cm$^{-2}$ per pulse). The $dR$ was measured under varying pump-probe time delay, meanwhile the energy $\omega$ of the pump and probe beams was degenerately swept across the exciton resonance, creating a 2D map of the $dR/R$ response as shown in Fig.~\ref{fig:one}(c). There are two main features to notice here. First is the change of $dR/R$ profile with increasing time delay: at short time delay the signal shows an antisymmetric form near the exciton resonance, similar to the reflectance contrast in Fig.~\ref{fig:one}(b) (also see line cuts in Fig.~\ref{fig:two}(a)); while, at longer time delay ($>5$~ps), the profile switches to a peak of predominantly positive signal which persists up to $>50$~ps. The second feature is the blue shift of the $dR=0$ position. If we simply attribute the $dR$ signals to the excitation induced bleaching (i.e., oscillator strength reduction), then the resulted $dR/R$ spectra should have an antisymmetric form similar to $\Delta R/R$ and the $dR=0$ position should not shift. Obviously the measured time dependence doesn't support such an assignment.

We note that our measured nonlinear response $dR/R$ can be viewed as the influence of the pump excitation to the exciton optical susceptibility $\chi$. Besides the oscillator strength reduction, the excitation can also lead to an energy shift of the exciton resonance through the many-body Coulomb interaction. We then write the nonlinear optical susceptibility as
\begin{equation}
\Delta\chi(\omega,\tau)=\frac{1-L(\omega)\Delta x(\tau)}{\omega-E_0-L(\omega)\Delta E(\tau)+i\gamma}-\chi_0.
\label{eq:NLsusc}
\end{equation}
Here $\chi_0=(\omega-E_0+i\gamma)^{-1}$ is the exciton optical susceptibility without the pump beam. The first term on the right hand side is the pump excitation modified susceptibility, with $\Delta x$ denoting the oscillator strength reduction and $\Delta E$ the exciton resonance shift. As the pump beam's energy is being degenerately swept across the resonance with the probe beam's, we used a standard Lorentzian shape $L(\omega)=((\omega-E_0)^2/\gamma^2+1)^{-1}$ to account for the $\omega$-dependence of the excitation population. $\Delta x$ and $\Delta E$ then only depend on the delay time $\tau$. The differential reflectance is related to the nonlinear susceptibility through $dR/R=A\cdot\mathrm{Re}(\Delta\chi)$, with the constant coefficient $A$ obtained from the antisymmetric Lorentzian fit to $\Delta R/R$ (Fig.~\ref{fig:one}(b)). By fitting the spectra at each delay (horizontal line cuts in Fig.~\ref{fig:one}(c)), the time dependences of both $\Delta x$ and $\Delta E$ are obtained. The resonance width $\gamma$ is chosen as the best fit to the full 2D data set. Note that in principle the pump beam can introduce a spectral broadening, i.e., in the right hand side of Eq. (\ref{eq:NLsusc}) the linewidth of the first term can be larger than the second term. We find its contribution to $\mathrm{Re}(\Delta\chi)$ is analogous to that from the oscillator strength reduction (both are antisymmetric, while the $\Delta E$ contribution is symmetric, see Fig. \ref{fig:one}(d)) thus its effect can be accounted by $\Delta x$. 

In Fig.~\ref{fig:two}(a) we plot several $dR/R$ cuts of constant delay time and their fits using Eq. (\ref{eq:NLsusc}). The fit performs remarkably well, capturing the important features of the data, namely the transition from an antisymmetric to symmetric profile and the blue shift of the $dR=0$ position.  By linearly interpolating in time between the individual fits of constant time slices a full 2D map of $dR/R$ can be simulated, as shown in Fig.~\ref{fig:two}(b), which matches up extremely well with the raw data of Fig.~\ref{fig:one}(c). The extracted time dependence of $\Delta x$ and $\Delta E$ are shown in Fig.~\ref{fig:two}(c) and (d), respectively. Both are found to exhibit bi-exponential decay forms:
\begin{align}
\Delta x(\tau)&=\Delta x_f e^{-\tau/\tau_{f}}+\Delta x_s e^{-\tau/\tau_{s}},\nonumber\\
\Delta E(\tau)&=\Delta E_+ e^{-\tau/\tau_+}-\Delta E_- e^{-\tau/\tau_-}.
\label{eq:biexponential}
\end{align}
Note that the $\Delta x$ parameter is composed of two positive terms. The $\Delta E$ parameter, however, is composed of a fast positive and a slow negative terms as clearly indicated in Fig. \ref{fig:two}(d). At short time delays, $\Delta E_+$ and $\Delta E_-$ largely cancel with each other and $\Delta x$ dominates the $dR/R$ responses; while at long time delays $\Delta E_-$ dominates. This leads to a transition from antisymmetric to symmetric profile with increasing time delay. The fits using Eq. (\ref{eq:biexponential}) are shown as the red curves in Fig. \ref{fig:two}(c) and (d). 

We now proceed to investigate the physical origin of $\Delta x_{f,s}$ and $\Delta E_\pm$. There are various physical mechanisms that contribute to the nonlinear optical response besides the well known phase space filling effect \cite{SuppleMater}. With a pump beam near resonant to the exciton energy it generates dense exciton populations, the exciton-exciton Coulomb interaction then brings a blue shift to the exciton resonance \cite{Rochat_PRB2000}. Besides the excitons, free carriers are also created by the pump excitation in the system. Generation of free carriers is reported even when the laser energy is significantly below the band-to-band transition \cite{Mitioglu_PRB2013,Pogna2016,Borzda2015}. Due to the strong Coulomb interaction in the 2D material, these free carriers can lead to a significant band gap renormalization which red shifts the exciton resonance \cite{Ugeda2014,Chernikov2015,Pogna2016,ChernikovPRL2015}, and also the plasma screening which not only decreases the exciton binding energy (resonance blue shift) but also reduces the oscillator strength. The net shift to the exciton resonance by the free carriers is usually negative \cite{SuppleMater}. Besides, both pump induced excitons and free carriers increase exciton scattering rate thus give rise to a spectral broadening, which in our model is accounted as a part of $\Delta x$ \cite{SuppleMater}.

From the above analysis, we assign the fast positive component $\Delta E_+$ to be the exciton-exciton interaction induced blue shift, this is further confirmed by its decay time $\tau_+\sim4$~ps which is consistent with previously reported exciton lifetimes \cite{Mai2014,Shi_ACSNano2013,Korn_APL2011,Lagarde2014a,GWang2014}. The slow negative component $-\Delta E_-$ is due to the combined effect from the free carrier band gap renormalization and plasma screening, with $\tau_-\sim40$~ps corresponding to the population decay lifetime of the free carriers. For $\Delta x$, its fast (slow) component $\Delta x_f$ ($\Delta x_s$) decays with a time scale $\tau_f\sim7$~ps ($\tau_s\sim70$~ps) similar to $\tau_+$ ($\tau_-$). Therefore we attribute $\Delta x_f$ to the phase space filling and scattering induced spectral broadening from the exciton population, while $\Delta x_s$ is the overall effect of the phase space filling, screening, and spectral broadening due to the free carrier population. The large difference between the two ratios $\Delta x_f/\Delta x_s\sim10$ and $\Delta E_+/\Delta E_-\sim1$ can be partly explained by the different scaling behaviors of $\Delta x_s$ and $\Delta E_-$ with the free carrier density $\rho_{eh}$. We expect $\Delta x_s$ to scale linearly with $\rho_{eh}$ while the exchange contribution to the band gap renormalization (in $\Delta E_-$) scales as $\sqrt{\rho_{eh}}$ \cite{SuppleMater}. Thus low density free carriers causes negligible oscillator strength reduction, but at the same time they can give rise to a significant resonance shift due to band gap renormalization. For $\Delta x_f$ and $\Delta E_+$, we expect they both scale linearly with the exciton density $\rho_X$ in the low density limit. From the extracted $\Delta E_+$ ($\Delta E_-$) value in Fig. 2(d), we roughly estimate that the peak exciton (free carrier) density is around the order of $\rho_X\sim10^{10}$~cm$^{-2}$ ($\rho_{eh}\sim10^7$~cm$^{-2}$) \cite{SuppleMater}.

To further support our interpretations we analyze the power dependent differential reflectance in a second sample. In the following measurements the probe power remains fixed at $15~\mu$W, while the pump power is varied from $5$ to $60~\mu$W.  For each power a full 2D map similar to Fig.~\ref{fig:one}(c) is created and fit at each delay time $\tau$ to extract out the power dependent time evolution of the $\Delta x$ and $\Delta E$ parameters as shown in Fig.~\ref{fig:four}(a) and (b), respectively (see the Supplemental Material \cite{SuppleMater} for the full power-dependent 2D data and the fits). We can see that the curves follow qualitatively similar patterns as Fig.~\ref{fig:two}(c) and (d), only the magnitudes change with the excitation power.

Each of the curves in Fig.~\ref{fig:four}(a) and (b) are then fit to a bi-exponential form (Eq. (\ref{eq:biexponential})), their best fit amplitudes ($\Delta x_f$, $\Delta x_s$, $\Delta E_+$ and $\Delta E_-$) are plotted in Fig.~\ref{fig:four}(c) and (d) and the corresponding decay times ($\tau_f$, $\tau_s$, $\tau_+$ and $\tau_-$) are plotted together in Fig.~\ref{fig:four}(e). In fitting $\Delta x$ under various pumping power, the power dependent amplitudes $\Delta x_f$ and $\Delta x_s$ are plotted in Fig.~\ref{fig:four}(c). $\Delta x_f$ (green dots), the oscillator strength change by the pump induced exciton population, shows a linear trend with increases pump power for low powers, and then saturates at higher powers. In contrast, $\Delta x_s$ (blue dots), the oscillator strength change by the pump induced free carrier population, is very weak at low pump power and grows superlinearly as the exciton population begins to saturate. Such behaviors suggest that the free carriers are created through the exciton Auger recombination process \cite{Zhu_SR2015,Kumar_PRB2014,Sun_NL2014,Mouri_PRB2014,Yu_EEA,Poellmann_NM2015}, where the collision between two excitons can nonradiatively annihilate one and transfer the energy to the second to ionize it to the free carrier continuum. Such a process leads to a quadratic dependence for the carrier density on the pump power.

\begin{figure}[t]
\includegraphics[width=1\linewidth]{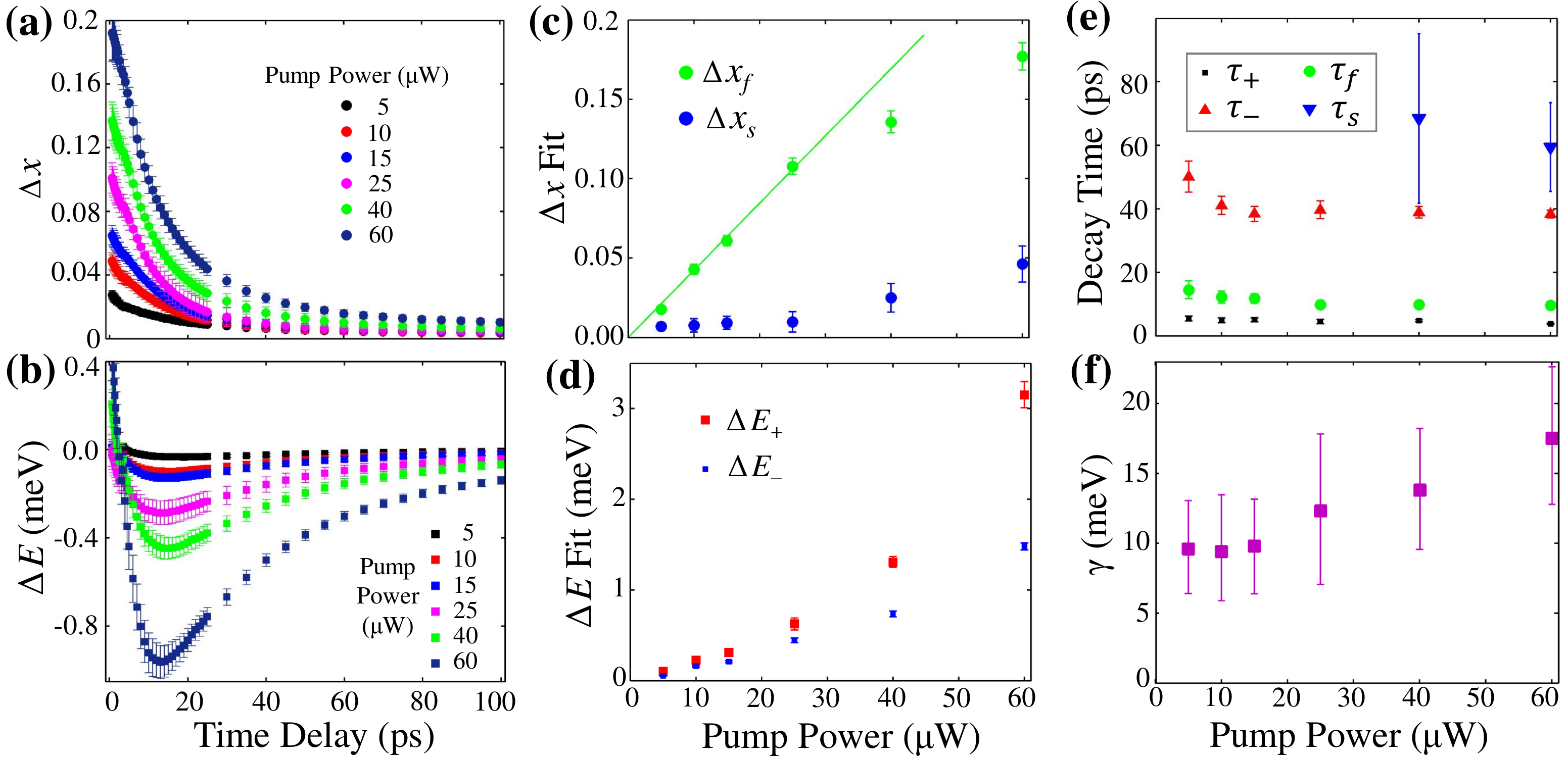}
\caption{\label{fig:four} 
Pump power dependence of fit parameters.  (a) and (b) plot the pump power dependent time evolutions of the the $\Delta x(\tau)$ and $\Delta E(\tau)$ fit parameters. (c) and (d) show the amplitude of the exponential temporal fits to the curves in (a) and (b) as a function of the pump power, respectively. Line is the linear fit to the low power data which deviates strongly to the high power data. (e) Plot of the decay time constants as functions of the pump power. The low power values of $\tau_s$ have very large errors thus are not shown. (f) Exciton resonance width increases with pump power.}
\end{figure}

Fig.~\ref{fig:four}(d) shows the magnitudes of the terms in the $\Delta E$ fit parameter which describes the effects of the exciton-exciton repulsion ($\Delta E_+$) and the free carrier induced resonance shift ($-\Delta E_-$). Under low power excitation $\Delta E_+$ (red dots) shows a linear trend with the pump power just as expected \cite{SuppleMater}. However, at higher power $\Delta E_+$ begins to grow superlinearly, implying that at high exciton densities the repulsion becomes stronger, and gives a nonlinear dependence on the exciton population. Similarly the $\Delta E_-$ term (blue dots) also shows a slight superlinear behavior under high power.

The power-dependence of the decay times ($\tau_f$, $\tau_s$, $\tau_+$ and $\tau_-$) are shown in Fig.~\ref{fig:four}(e) as a function of the pump power. In general all decay times show a decrease with the increasing power. The two terms due to the exciton population ($\tau_+$ and $\tau_f$, black and green dots, respectively) are in general much faster than the terms due the the free carrier population ($\tau_-$ and $\tau_s$, red and blue dots, respectively) meaning that the optically active excitons decay faster than the optically inactive free carriers. Additionally the measured exciton decay timescales of $3-10$~ps again agree well with the previously reported values for the exciton lifetimes \cite{Mai2014,Shi_ACSNano2013,Korn_APL2011,Lagarde2014a,GWang2014}.

Finally we note that the width of the resonance, $\gamma$, which is derived as a global best fit independent of the delay time $\tau$ to each 2D map, generally increases with the pump power as shown in Fig.~\ref{fig:four}(f). This is consistent with the decreasing lifetimes with the increasing power as shown in Fig.~\ref{fig:four}(e), larger pump induced populations lead to stronger interactions and hence a decrease in the overall coherence lifetime, which manifests as an increase in the resonance width \cite{Moody_NC2015}. But when the pump power is below the probe power (15 $\mu$W), $\gamma$ is nearly unchanged. This could be a signature of probe induced linewidth broadening \cite{SuppleMater}.

These observed significant many-body effects on exciton resonance in monolayer WSe$_2$ confirms the important role of the Coulomb interaction in 2D TMDs. Importantly, due to the short lifetime of the excitons, it is the free carrier band gap renormalization effect that dominates the long time behaviors of the nonlinear optical responses. Further understanding of these effects can shed light on the interesting physics in 2D semiconductors.

\begin{acknowledgements}
The work at UW is mainly supported by the U.S. Department of Energy (DOE), Basic Energy Sciences (BES), Materials Sciences and Engineering Division (DE-SC0008145, DE-SC0012509, and DE-SC0002197). H.Y. and W.Y. were supported by the Croucher Foundation (Croucher Innovation Award), the RGC and UGC of Hong Kong (HKU17305914P, AoE/P-04/08), and the HKU ORA. J.Y. and D.G.M. were supported by the DOE, BES, Materials Sciences and Engineering Division. X.X. acknowledges a Cottrell Scholar Award. 
\end{acknowledgements}

\end{document}